# Bach2Bach: Generating Music Using A Deep Reinforcement Learning Approach

Nikhil Kotecha
*Columbia University*

**Abstract**

*A model of music needs to have the ability to recall past details and have a clear, coherent understanding of musical structure. Detailed in the paper is a deep reinforcement learning architecture that predicts and generates polyphonic music aligned with musical rules. The probabilistic model presented is a Bi-axial LSTM trained with a "kernel" reminiscent of a convolutional kernel. To encourage exploration and impose greater global coherence on the generated music, a deep reinforcement learning (DQN) approach is adopted. When analyzed quantitatively and qualitatively, this approach performs well in composing polyphonic music.*

## 1. Introduction

This paper describes an algorithmic approach to the generation of music. The key goal is to model and learn musical styles, then generate new musical content. This is challenging to model because it requires the function to be able to recall past information to project in the future. Further, the model has to learn the original subject and transform it. This first condition of memory is a necessary and non-trivial task. The second necessary condition is cohesion: the next challenge is to understand the underlying substructure of the piece so that it performs the piece cohesively. It is easier to create small, non-connected subunits that do not contribute to sense of a coherent piece. A third non-necessary, but sufficient condition is musical, aesthetic value and novelty. The model should not be recycling learned content in a static, thoughtless way. Rather, the model should optimize under uncertainty, exploring the potential options for note sequences and select the highest valued path.

In the following subsections of the introduction, some musical language will be introduced, followed by the necessary conditions and then the sufficient conditions. An outline of solutions will be also provided. In this paper, two different papers will be re-implemented and used as benchmarks, changes will be described. In subsequent sections, the following will be covered: methodology, objectives and technical challenges, problem formulation and design, implementation, results and conclusion.

## 1.1 Music as a sequence generation task

One method to algorithmically generate music is to train a probabilistic model. Model the music as a probability distribution, mapping measures, or sequences of notes based on likelihood of appearance in the corpus of training music. These probabilities are learnt from the input data without prior specification of particular musical rules. The algorithm uncovers patterns from the music alone. After the model is trained, new music is generated in sequences. This generated music comes from a sampling of the learned probability distribution. This approach is complicated by the structure of music. Structurally, most music contains a melody, or a key sequence of notes with a single instrument or vocal theme. This melody can be monodic, meaning at most one note per time step. The melody can also be polyphonic, meaning greater than one note per time step[2]. In the case of Bach's chorales, they have a polyphony, or multiple voices producing a polyphonic melody. These melodies can also have an accompaniment. This can be counterpoint, composed of one or more melodies or voices[3]. A form of accompaniment can also be a sequence of chords that provide an associated term called a harmony. The input has great bearing on the nature of the output generated.

These musical details are relevant because training a probabilistic model is complicated by the multidimensionality of polyphonic music. For instance, within a single time step multiple notes can occur creating harmonic intervals. These notes can also be patterns across multiple time steps in sequence. Further, musical notes are expressed by octave, or by interval between musical pitches. Pitches one or more octaves apart are by assumption musically equivalent, creating the idea of pitch circularity. Pitch is therefore viewed as having two dimensions: height, which refers to the absolute physical frequency of the note (e.g. 440 Hz); and pitch class, which refers to relative position within the octave. Therefore, when music is moved up or down a key the absolute frequency of the note is different but the fundamental linkages between notes is preserved. This is a necessary feature of a model. Chen et al[4] offered an early paper on deep learning generated music with a limited macro structure to the entire piece. The model created small, non-connected subunits that did not contribute to a sense of a coherent composition. To effectively model music, attention needs to be paid to the structure of the music.

**1.2 Necessary Conditions: Memory and Cohesion**

A model of music needs to have the ability to recall past details and understand the underlying sub-structure to create a coherent piece in line with musical structure. Recurrent neural networks (RNN), and in particular long short-term memory networks (LSTM), are successful in capturing patterns occurring over time. To capture the complexity of musical structure vis a vis harmonic and melodic structure, notes at each time step should be modeled as a joint probability distribution. Musically speaking, there should be an understanding of time signature, or the number of notes in a measure. Further, the RNN-based architecture allows the network to generate notes identical for each time step indefinitely, making the songs time-invariant. Prior work with music generation using deep learning [5] [6] have used RNNs to learn to predict the next note in a monophonic melody with success.

To account for octaves and pitch circularity, greater context is needed. Following the convolutional neural network (CNN) architecture, a solution is to employ a kernel or a window of notes and sliding that kernel or convolving across surrounding notes. Convolutional layers have a rich history linked to the mammalian visual system. For instance, "a deep convolutional network… uses hierarchical layers of tiled convolutional filters to mimic the effects of receptive fields occurring in early visual cortical development[7]." Much in the same way vision requires invariance in multiple dimensions, CNNs offer a way to develop hierarchical representations of features giving invariance to a network. In effect, this allows the model to take advantage of "local spatial correlations" between measures, and building in robustness to natural transformations. Musically speaking, notes can be transposed up and down staying fundamentally the same. To account for this pitch circularity, the network should be roughly identical for each note. Further, multiple notes can be played simultaneously - the idea of polyphony - and the network should account for the selection of coherent chords. RNNs are invariant in time, but are not invariant in note. A specific output node represents each note. Moving, therefore, up a whole step produces a different output. For music, relative relationships not absolute relationships are key. E major sounds more like a G major than an E minor chord, despite the fact that E minor has closer absolute similarity in terms of note position. Convolutional networks offer this necessary invariance across multiple dimensions. Inspired by Daniel Johnson's[8] Bi-axial LSTM model, I describe a neural network architecture that generates music. The probabilistic model described is a stacked recurrent network with a structure employing a convolution-esque kernel.

The model described thus far learns past information so that it can project in the future. The solution described is to use a LSTM network. There are limitations to using an RNN model. As stated initially, an essential feature of an algorithmic approach to music is an understanding of the underlying substructure of the piece so that it performs the piece cohesively. This necessitates remembering past details and creating global coherence. An RNN solves this problem generally: by design an RNN generates the next note by sampling from the model's output distribution, producing the next note. However, this form of model suffers from excessive repetition of the same note, or produces sequences of notes that lack global coherent structure. The work can therefore sound meandering or without general pattern.

**1.4 Sufficient Condition: Musical Novelty**

The next challenge is to understand the underlying substructure so that it performs cohesively. The convolutional-esque kernel offers greater context needed for musical coherence, but is insufficient. A third non-necessary, but sufficient condition of generating music is aesthetic value, increased coherence, and novelty. This third condition is difficult to model due to the subjective nature of what makes a song sound "good." A way to solve a related problem is to allow for exploration. Instead of sampling from a static learned distribution as in the case of a pure deep learning approach, the reinforcement learning (RL) algorithm can be cast as a class of Markov Decision Processes (MDP). An MDP is a decision making framework in which outcomes are partly random and partly under the control of a decision maker. (More precise details on reinforcement learning and MDPs in the methodology section.) At each time step, an agent can visit a finite number of states. From every state there are subsequent states that can be reached by means of actions. When a state is visited, a reward is collected. Positive rewards represent gain and negative rewards represent punishment. The value of a given state is the averaged future reward which can be accumulated by selecting actions from the particular state. Actions are selected according to a policy which can also change. The goal of an RL algorithm is to select actions that maximize the expected cumulative reward (the return) of the agent. The approach will be described in greater detail in the methodology section.

In the context of music, the necessary and sufficient conditions described above combine to create a sequence learning and generation task. RL is used to impose structure on an Bi-axial LSTM with a covolutional-esque kernel trained on data. The reward function is a combination of rewards associated with following hard-coded musical theory rules and a reward associated with the probability of a given action learned by the LSTM network. This enables an accurate representation of the source probability distribution learned from Bach's music, while still retaining musical constructs - pitch, harmony, etc - to bound the samples within reasonable, heuristic musical rules. This mix of reward learned from data and task specific reward combined into a general reward function provides a better metric tailored to the specific task of generating music. Different from previous approaches [9][10][11][12] and following the lead of [13], the model mainly relies on information learned from data with the RL component improving the structure of the output through the imposition of musical, structural rules.

Overall inspired by Daniel Johnson's[8] Bi-axial LSTM model and Natasha Jacques'[13] Reinforcement Learning model, I describe a deep neural network with reinforcement learning architecture that generates music. The probabilistic model described is a stacked recurrent network with a structure employing a convolution-esque kernel, refined by a RL component. Presented is the model, the approach to training, and generation.

## 2. Background

### 2.1 Deep Q-Learning

A song can be discretized and interpreted as a series of finite measures of notes concatenated into a full song. Given the state of the environment at time $t$, $s_t$, the agent takes an action according to its policy $(a_t|s_t)$, receives a reward $r(s_t, a_t)$ and the environment transitions to a new state, $s_{t+1}$. The agent's goal is to maximize reward over a sequence of actions, with a discount factor of $\gamma$ applied to future rewards. Casting the problem in this framework, yields traction from a reinforcement learning and dynamic programming approach.

A dynamic programming and RL approach splits the multi-period planning problem, as in the case of music, into easier sub-problems at different points in time. Information describing the evolution of the decision problem over time is therefore necessary. The LSTM approach solves this problem and encodes this information in the forget and input gate mechanism (described further in the subsequent background section). The information necessary about the current situation needed to make the "correct" decision, that which maximizes the expected reward is achieved through an RL or dynamic programming approach. In general, RL methods are used to solve two related problems: Prediction Problems and Control Problems. In prediction problems, RL is used to learn the value function for the policy followed. At the end of learning, the learned value function describes for every visited state how much future reward can be expected when performing actions starting at this state. Control problems takes this a step further. Interaction with the environment offers a chance to find a policy that maximizes reward.

By traveling through state space, the agent is learning the optimal policy[14]: a rule that determines a decision given the available information in the current state. After sufficient traveling, the agent obtains an optimal policy which allows for planning of actions and optimal control. If the control problem is reframed as a predictive type of control, the solution to the control problem appears to require a solution to the prediction problem as well.

From Richard Bellman [15], an optimal policy has the property that whatever the initial state and initial decision are, the remaining decisions must constitute an optimal policy with regard to the state resulting from the first decision. The problems that can be broken apart like this have in the world of computer science "optimal substructure", which is analogous to the idea of "subgame perfect equilibria" from game theory. (Notation used is consistent with Jacques paper.) The optimal deterministic policy $\pi^*$ is known to satisfy the Bellman optimality equation [15]:

$$(1) \quad Q(s_t, a_t; \pi^*) = r(s_t, a_t) + \gamma \mathbb{E}_{p(s_{t+1}|s_t,a_t)}[max_{a_{t+1}} Q(s_{t+1}, a_{t+1}; \pi^*)]$$

Where $Q^\pi(s_t, a_t) = \mathbb{E}[\Sigma_{t'=t}^\infty \gamma^{t'-t} r(s_{t'}, a_{t'})]$ is the Q function of a policy $\pi$. The Bellman equation shows that a dynamic optimization problem in discrete time can be expressed recursively by relating the value function in one period relative to the next. The optimal policy in the last time period is specified in advance as a function of the state variable's value at that time. The following optimal value objective function can be then expressed in terms of that state variable. This continues, maximizing the sum of the period's time specific objective function. Using recursion, the first period decision rule can be derived as a function of the initial state variable value by optimizing the sum of the first period specific objective functions and the one step look ahead, which captures the value for all future periods. Therefore, each period's decision is made by acknowledging that all future decisions will be optimally made. Practically, since Bellman's equation is a functional equation solving the Bellman equation solves for the unknown value function. The value function is a function of the state and characterizes the best possible value of the objective. By calculating the value function, the function describing the optimal action as a function of the state is also found, called the policy function.

This is useful in the context of music generation because the Bellman equation offers a method to solve stochastic optimal control problems, like a Markov Decision Process. A Markov Decision Process (MDP) is a discrete time stochastic control process. At each time step, the process is in some state, and the decision maker may choose an action in said state. The process responds by transitioning to a new state and giving the decision maker a corresponding reward. The probability that the process moves into its new state is influenced by the agent's chosen action, characterized by the state transition function. Given the state and action, state transitions are conditionally independent of all previous states and actions. The Markov property holds: recent past, not distant past influence the next state. The central goal of MDPs is to find a policy function, or the action that the agent takes given the state. The desire is to select a policy that maximizes the cumulative function of the random rewards, usually the expected discounted sum over a potentially infinite horizon (termed an infinite horizon MDP). This entails a discount factor multiplied the reward which is a function of the state summed over the infinite horizon. A MDP can also be thought of as a one-player stochastic game[16]. If the probabilities or rewards are unknown, this becomes a reinforcement learning problem.

At a high level, there exist several different ways of finding the optimal value function and/or the optimal policy. If the state transition function $T(s, a, s')$, which characterizes the transition probability in going from state $s$ to $s'$ when performing action $a$. And if the reward function $r(s,a)$ which determines how much reward is obtained at a state, then algorithms which can be modeled are called model-based algorithms. They can be used to acquire the optimal value function and/or the optimal policy. Of note are value iteration and policy iteration, which both come from Dynamic Programming [15], not RL. These two approaches are beyond the scope of the methodology, but the names are included for completeness. If the model of the process, namely the transition function and the reward function are unknown ex ante, then this becomes a RL problem. In the language of control theory, an adaptive process of the optimal value function and/or the optimal policy will need to be learned. Notable algorithms include: temporal difference learning (TD) which in isolation is used for value function learning; Adaptive Actor-Critics, which is an adaptive policy iteration algorithm used to approximate the model of the value function by TD where the TD error is used for the actor and the critic; and most relevant for this paper is Q-learning, which allows for concurrent value function and policy optimization.

In Q learning, an action is taken, and given uncertainty over the transition probabilities or rewards the agent continues optimally given the current policy. Experience during learning follows: given the current state and the taken action, a new state emerges. Q-learning techniques [17] [18] learn this optimal Q function by iteratively minimizing the Bellman residual. The optimal policy is given by: $\pi^*(a|s) = argmax_a Q(s,a)$. Deep Q-learning [19] uses a neural network called the deep Q-network (DQN) to approximate the Q function, $Q(s,a;\theta)$. This naive approach has some major flaws, namely the Q function can diverge when a non-linear function approximator, such as a neural network is used [20]. Solutions proposed by Mnih et al.

[19] use a method termed experience replay that "randomizes over data, removing correlations in the observation sequence and smoothing over changes in the data distribution." Mnih et al also propose an iterative update that adjusts the action values toward target values that are only periodically updated. In effect, the network parameters $\theta$ are learned by applying stochastic gradient descent (SGD) updates with respect to the following loss function,

$$(2) \quad L(\theta) = \mathbb{E}_\beta[r(s,a) + \gamma max_{a'} Q(s', a'; \theta^-) - Q(s, a, \theta))^2]$$

where $\beta$ is the exploration policy, and $\theta^-$ is the parameter of the Target Q-network [19] that is held fixed during the gradient computation. The moving average of $\theta$ is used as $\theta^-$ as proposed in [21]. Exploration can be performed with either the epsilon-greedy method or Boltzmann sampling. Additional standard techniques such as replay memory [20] mentioned above and Deep Double Q-learning [22] are used to stabilize and improve learning. In game theoretic language, the agent is exploring the potential sub-game equilibria and finding the corresponding policy functions to approximately solve the infinite horizon MDP through the Bellman equation, practically through the DQN.

**2.2 LSTMs**

Recurrent networks encounter a serious problem caused by difficulty in estimating gradients. In backpropagation through time (BPTT), recurrence passes multiplications in repetition. This can lead to diminishingly small or increasingly large effects, respectively called the vanishing or exploding gradient problem. To resolve this problem, Hochreiter and Schmidhuber[23] designed Long short-term memory (LSTM) networks. The LSTM is designed to secure information in memory cells, separate and protected from the standard information flow of a recurrent network. To pass, read or forget information is performed by opening or closing the input gate, output gate, or forget gate. This process of is akin to a neuron firing. Input gates and output gates control flows of information in and out of the cell. The forget gate controls if information in the cell should be reset. LSTMs are better at learning long-term dependencies in data, and readjust to data in a fast manner[24]. LSTMs are an effective combination with the softmax function. A softmax function, a generalization of a logistic function, squashes arbitrary real values to values in the range (0,1) and that add up to 1. This property makes the softmax function effective at representing a categorical distribution. In music generation, the softmax function takes as input the network output of the LSTM and outputs probability values assigned to different notes available to play. LSTM gates are modulated by weight that is differentiable, allowing for backpropagation through time (BPTT). Softmax cross-entropy loss is used to train the model in typical neural network learning fashion, through BPTT [25]. To reiterate from a previous section, songs generated using an only deep learning approach lack global structure. An RL approach can improve this model.

## 2. Methodology

In this section, presented is Daniel's Johnson's original model followed by extensions to the model. In the original paper there are a few models attempted to generate music. Here the best performing model is selected, replicated, and the model is extended. Additionally, presented is Natasha Jacques et. al's deep reinforcement learning approach primed with my extended model.

### 2.1. Objectives and Technical Challenges

One key challenge with modeling music is selecting the data representation. Possible representations are signal, transformed signal, MIDI, text, etc. In general, musical content for computers is first represented as an audio signal. It can be raw audio (waveform), or an audio spectrum processed as a Fourier transform. A relevant issue is the end destination of the generated music content[2]. The format destination could be a human user, in which case the output would need to be human readable, for instance a musical score. In the case of this paper, the destination is a computer. The final output format is therefore readable by a computer, which in this case is a MIDI file (musical instrument digital interface). The MIDI representation was selected because it offers a particularly rich representation in two senses: first it carries characteristics of the music in the metadata of the file, like time steps. Second it is a common digital representation which allowed access to freely and widely available data. In this model, the criteria optimized for are: computer readable, information about characteristics of the music, and availability to a wide selection of Bach's work. More detail will be provided about each of these choices.

    The choice of the MIDI file format has substantial bearing on the model. There is a question of how much richness to have in the objective characteristics of the music, or the sense of musical structure. Richer musical representation affords greater precision in the potential playing, but it also creates a more supervised approach to the generation. One must for instance know the musical theory in a deep way to understand the different characteristics of a musical score. The level of musical detail beyond the waveform therefore represents a choice of how much or how little to include. As stated above, the MIDI file format offered the greatest balance between objective characteristics of the music and the raw waveform and was thus selected for training. Other considered data sets are worth mentioning because they offer consideration for different definitions for the "generation" of music.

One data set is called Bach Digital [26]. The population covered is 90 percent of Bach's compositions in high resolution scans of his work. The collection data is variable, but in some ways is irrelevant because the scores were produced several hundred years ago. The topics covered are a major portion of Bach's oeuvre. This is a useful dataset because it has the clearest representation of how Bach wanted his music. Granted there is a lot of entropy in interpretation, but this is as close to the man as we can get. In the sense of objective music characteristics (accents, fermatas, loudness, etc), this is the optimal data set. The challenge with this dataset is getting the information in a computer readable format. In many ways it is easier to go from the raw waveform and have the machine learn from the sound straight. But the loss of fundamental musical information is significant. The midi format (dataset selected) offers a hybrid of waveform and objective characteristics but still it is a crude representation of the music. The challenge of converting the musical score to a waveform is a non-trivial task beyond the scope of this paper. The deeper principle here is the level of supervision in the generation of the output. At one extreme is complete autonomy and automation with no human supervision. Or it could be more interactive, with early stopping built into the model to supervise the music creation process. The pure neural network approach employed by this paper is by design non-interactive. The MIDI file format optimized for this dimension as well because it offers a complete end product that is machine readable without human intervention. The level of autonomy is an interesting potential development for actual musicians who can interrupt the model in the middle of content generation: have a human guide the process of generation. While beyond the scope of the paper, feedback throughout the process can lead to suggestions that are superior, or more aesthetically pleasing musical compositions.

Another data set is called Bach Choral Harmony data set [27]. The population is composed of 60 chorales by Bach. The dataset donated in 2014 comes from a free resource called the UCI machine learning repository. This data set is useful because it offers a textual representation of Bach's work. It contains pitch classes extracted from midi sources, meter information, and chord labels (human annotated). In this way it offers an additional pairing to the raw waveform. The challenge with this dataset is if it is used in isolation. It is tough to go from the text to an audio representation, and therefore as such does not provide a closed form solution to the original posed question. This data set offers an interesting addition to the other data sets because it gives information that Bach digital data set has encoded, information straight from the scores. These objective characteristics are valuable because they give a sense of what Bach himself was trying to create. Performing a pattern recognition on the source limits the potential loss of information. Some general information on how textual representation works. A melody can be processed as text. One common example in folk music is the ABC notation. The first 6 lines provide metadata – e.g. T: title of the music, M: Meter (time signature is a more accurate term), L: default length, K: Key, etc. This is then followed by the main text which codifies the melody. A key thing captured is the pitch class of a note is encoded by association with its English notation. The pitch is represented by with either an upper or lower case. Upper case A for instance represents A above middle C (A440 Stuttgart Pitch). a represents one octave up and a' represents two octaves up. A further consideration with this data set is the focus on Bach's chorales. The specific choice of what selection of music to train on is a judgement that has great bearing on the resulting output. Interpretation is a significant source of artistry and generation. The breadth of the sample offers a wide potential for probability distributions. The breadth of the sample also limits the ability to bore deeply into one song. If instead of a wide slice, I selected a data set with one song – for instance, a personal preference of Bach's music Partitia no. 2 in D minor – but with several artists I would have a learned manifold with a rich and deep understanding of one piece. Understanding how Joshua Heifetz, Itzhak Pearlman, Ivry Gitlis, and say Hilary Hahn all play the same song differently could yield great depth of understanding, a complex manifold in its own right. If Bach's oeuvre is the state space for the model, the selection of which songs and how many songs seems to suggest a generalist versus specialist tradeoff in the output of the model. In the training of this model I selected a more generalist approach.

The last dataset considered is Bach's sheet music [28]. This is misleading because the listed scores are actually lead sheets (mostly). The population covered is a significant portion of Bach's violin compositions. Lead sheets are an important representation because they convey in a single or few pages the key ideas of a piece: the score of a melody with annotations specifying harmony (chord labels). Also given are composer, musical style (e.g. detache, legato, staccato), and tempo (allegro for instance). The salient details are given in a data rich and concise format easy to add to the music directly (by hand). For edification, lead sheets are often used in Blues for improvisation (where my familiarity comes from). This data set is by design a form of lossy compression. It takes the data set and represents it in a less memory intensive, inexact approximation, partial data format. Therefore there is a lot of information lost in this form. This is good for the purposes of adding a few objective characteristics but is an inefficient estimator because it fails to capture the full source of Bach's music.

As stated originally, the MIDI file format was selected for its balance of musical information with raw waveform, a wide availability of Bach's oeuvre, and its computer readable format. Next the model details will be explored.

**2.2. Problem Formulation and Design**
**2.2a Deep Learning Network**

To capture the harmonic and melodic structure between notes, the model uses a two-layered LSTM RNN architecture with recurrent connections along the note axis. By having one LSTM on the time axis and another on the note axis, the model takes on, to borrow Daniel Johnson's language[8]: a "bi-axial" configuration.

The note-axis LSTM receives as input a concatenation of final output of the note-axis LSTM for the previous note window and the activations of the last time-axis LSTM layer for the particular note. The output of the final activations of the note-axis LSTM are then fed into a softmax layer to convert to a probability. The loss corresponds to the cross entropy error of the predictions at each time step compared to the played note at each time step. Each note therefore has a time component from the time-axis LSTM. This allows for understanding the temporal relationships for the particular note and for modeling the joint distribution of notes in the particular time step. By joining the information from an LSTM focused on the time component and an LSTM focused on the note-component, the relationships within and between notes is captured for each timestep. By using this approach on each note in sequence, the full conditional distribution for each time step can be learned. Further, another key piece of functionality is building into the model a window that slides over sequences of notes. This architecture enables the model to learn the harmonic and melodic structure of the notes accounting for pitch circularity.

Extending beyond Daniel Johnson's model, the model presented here is designed and implemented to be flexible, general, and to take advantage of parallelization in code. A primary goal was flexibility in user input. The architecture is general: the user can set various hyper parameters, such as the number of layers, hidden unit size, sequence length, time steps, batch size, optimization method, and learning rate. The model is parameterized so users can also set the length of the window of notes fed into the note-axis LSTM model and the length of time steps fed into the time-axis LSTM. The size of the window and length of the time steps are a relevant features because music is highly variable based on genre and artist. Designing the system to be general allows the user to tailor the model to his/her specific needs. In terms of functionality and model design, a primary goal for the model was parallelization in code. The code was written at a high level to do everything in efficient matrix format, minimizing the use of 'for' loops. This allows for speed gains in computational time.

### 2.2b Reinforcement Learning Framework

With the trained modified Bi-axial model described above (henceforth "Biaxial model"), the next part of the process is to have the model learn musical theory concepts. To achieve this, I follow Natasha Jacques et al.'s model [13] of a RL tuner. The LSTM trained on data (the Biaxial model) primes the three networks in the RL model: the Q-network and Target Q-network in the DQN algorithm described in section 2.1, and a Biaxial Reward. The Q-network is initialized with an LSTM model with architecture identical to that of the Biaxial model. The Biaxial Reward furnishes part of the reward value used to train the model and is fixed while training. In order to cast musical sequence generation in the RL framework, each subsequent note in the sequence or melody is seen as taking an action. The state, $s$, is the previous note and the internal state of the LSTM cells of both the Q-network and the Biaxial Reward. Q(a,s) can be calculated by initializing the Q-network with the suitable memory cell contents, running it one time step using the previous note, and evaluating the output value for the action a. The next action can be selected with either a Boltzmann sampling or epsilon-greedy exploration strategy.

A key point is whether RL can be used to offer bounds on a sequence learner such that the sequences it generates conform to a desired, specified structure. To test this idea, I codified musical rules consistent with some notions from texts on musical composition [29] [30] [31] [32] [33]. These musical rules are heuristic: I am not a professional musician and as such have chosen rules with what I thought most salient. The goal of these rules is to steer the model toward more traditional melodic composition. The codified musical rules are a sufficient addition to the necessary Biaxial LSTM model, which learns directly from the data.

## 3. Implementation

In this section, the process of training the network and the generation of new musical compositions will be explained. Experiments were performed on Google Cloud Platform with deep learning implementation done in TensorFlow. Sources of material that helped guide the implementation: Daniel Johnson's code [34]. For loading the data into the appropriate format [35].

### 3.1. Deep Learning Network

The model is applied to a polyphonic music prediction task. The network is trained to model the conditional probability distribution of the notes played in a given time step, conditioned on the notes in previous time steps. The output of the network can be read as at time step t, the probability of playing a note at time step t, conditioned on prior note choices. Therefore, the model is maximizing the log-likelihood of each training sequence under the conditional distribution.

The time-axis LSTM depends on chosen notes, not on the specific output of the note axis layers. The rationale is that all notes at all timesteps are known so training can be expedited. The time gain comes from processing the input, then feeding the pre-processed input through the LSTM time-axis in parallel for all notes. Next, the LSTM note-axis layer computes the probabilities across all time steps. This provides a significant speed up when using a GPU to perform parallel computing.

Now that the probability distribution is learned, sampling from this distribution offers a way to generate new sequences. Sequences are not known in advance. The network must project one time step in the future at a time. The input for each timestep is used to advance the LSTM time-axis layers one step at at a time to compose the note in the next period. First a sample must be taken while the distribution is being created. Each note is drawn from a Bernoulli distribution. This drawn value is then used for the input to the next note. This process is repeated for all notes, after which the model moves to the next time step.

The model was tested on a selection of Bach's works from [17] as well as the classical piano files from [18]. Input was in the form of MIDI files.

After training the Bi-axial LSTM, the model was used to create new musical compositions. A larger and diverse dataset with different note and structural patterns was used during training. The goal here was to expose the model during training to a wide variety of patterns so as to encourage as much diversity in output as possible. The MIDI file format enables the use of a temporal position in the music. A time component was an important feature to build into the dataset so that the model could learn patterns over time relative to different note sequences. Following the guide of Johnson[8], an additional dimension was added to the note vectors fed into the model: a binary, 0 or 1 to indicate if a note was articulated or sustained at a particular time step. From Johnson, for instance, the first time step for playing a note is represented as 11. Sustaining a previous note is represented as 10, and resting is represented as 00. This added dimension allows the model to play the same note multiple times in succession. From the input perspective, the articulation dimension or bit is processed beforehand. This processing is done in parallel with the playing dimension, which together are then fed into to the time-axis LSTM. From the output perspective, the note-axis LSTM gives a probability of playing a note and a probability of articulating the same note. When computing the cost function, articulating a played note incorrectly is penalized. The articulation output for notes that should not be played is ignored. It makes little sense to penalize for articulation if a note is not played.

Using Moon et al[38] as a suggested guide, Dropout of .75 was applied to each LSTM layer. The optimizer selected was ADADELTA[39]. The learning rate selected was 1.0. The Biaxial models were evaluated in two dimensions.

### 3.2 Reinforcement Learning

Following from section 2.2b, music generation can be cast as an RL problem if placement of the next note in the song is treated like an action. The state is the environment consisting of the previous note and the internal state of the Q network and the Biaxial Reward. Given the action, reward can be evaluated by joining the Biaxial model's prediction probability of the accurate note learned from the data guided with the hard coded musical theory rules. The reward, in short, can be seen as a combination of the best predicted note as learned from data and from musical theory rules. Described at a high level in section 2.2b, musical theory rules are defined to offer bounds on the song that the model is composing through the reward $r_{MT}(a,s)$. If a note, for instance, is in the wrong key, then the model will receive a punishment or a negative reward. There is a tradeoff present in this decision to incorporate a model learned from data and a model constrained exogenously: more constraint or tighter bounds offer narrower searching of the state space resulting in more similar actions and more uniform melodic composition, but a consistent sounding melody. On the other side, fewer constraints offer greater melodic "creativity" in the sense that the model will move beyond creating a simple melody that exploits the sure-fire rewards and satisfy the initial sufficient condition of novelty. To accomplish this, the Biaxial Reward is used to compute $logp(a|s)$, the log probability of a note $a$ given a melody $s$, and incorporate this into the reward function.

The total reward given at time t is:

$$r(s,a) = logp(a|s) + r_{MT}(a,s)/c \quad (3)$$

The constant c controls the weight given to the musical theory. From the DQN loss function in equation 2 and the above reward function in equation 3, the modified loss function and learned policy are as follows:

$$L(\theta) = \mathbb{E}_\beta[(logp(a|s) + r_{MT}(a,s)/c + \gamma max_{a'} Q(s',a';\theta^-) - Q(s,a;\theta))^2] \quad (4)$$

$$\pi\theta(a|s) = \delta(a = argmax_a Q(s,a;\theta)) \quad (5)$$

The loss function encourages the model to value actions in accordance with the specified musical rules and from the source material, selecting notes with high probability of matching the learned distribution.

To understand the specific musical rules that should be encoded into the reward, I read several music theory books [29] [30] [31] [32] [33]. I tried to characterize which principles were preserved in multiple books, most relevant to general composition, agreed with the described rules in Jacques et al's paper, and were mathematically interpretable. The music reward function $r_{MT}(a, s)$ was foster the following characteristics. Generated notes should belong to the same key, with the melody starting and finishing with same tonic note of the key. For instance, if the key is in D-major, the rewarded note would be a middle D. The produced note should also occur in the first beat and the last four beats of the melody. There should be a designation if a rest is introduced or a note is held (see discussion above in section 3.1, last paragraph on articulation). Outside the cases of a rest or a held note, repetition of the same sound should be minimized. Specifically, a single tone should not be repeated more than four times in a row. These specifications offer bounds on the generated music. To satisfy the sufficient condition of novelty, the model receives a negative reward if the model is too similar with itself from previous time steps. Repeated patterns can be identified by looking at similarity between observations as a function of time lag, termed autocorrelation. Practically, autocorrelation measures the correlation of a signal and its copy as a function of time. To prevent excessive local periodicity, autocorrelation is used to look at time domain signals. Signal processing has a rich history of using autocorrelation [40]. In the context of music, this can be done by measuring the autocorrelation at a time lag of one, two, or three beats. Specifically, the negative reward is applied when the autocorrelation coefficient is larger than .15. Further, the melody should follow traditional musical intervals. The size of an interval between two notes is the ratio of their musical frequencies. The size of the main intervals can be expressed by integer ratios. For instance, some key intervals: one-to-one is called the unison; two-to-one is called an octave; three-to-two is called the perfect fifth; four-to-three is called the perfect fourth; five-to-four is called the major third; six-to-five is called the minor third. Pitch increments following the same interval produce an exponential increase of frequency, despite the fact that people physically perceive this increase as a linear increase in pitch. This results in the fascinating idea that people perceive logarithmically [41]. An interval can also be described as horizontal consonance or dissonance, commonly called melodic if successive or prior notes in a temporal sense have adjacent pitches (right to left, left to right). An interval can also be thought of in a vertical consonance or dissonance sense, commonly called harmonics if concurrent notes have adjacent pitches while interacting in a spatial sense (up and down, down and up). Melodies should avoid non-traditional intervals, intervals occuring outside the described increments. As an example, the chord made up two major thirds, called an augmented triad should fall within traditional increments avoiding clumsy intervals like augmented sevenths. Large jumps of greater than an octave should also receive a negative reward. Gauldin [29] and Rimsky-Korsakov [32] recommend good composition balance continuity and novelty: most often moving in small steps and occasionally large harmonic intervals. The reward function aligns with this rule. Following Schoenberg's Fundamentals of Musical Composition [31], melodic movement should follow a sense of mean reversion. If the melody moves a fifth or more in a direction, actions that return the agent with a leap or a gradual movement in the opposite direction are rewarded. Significant melodic movement in the same direction is disincentivized. highest and lowest notes should not be repeated. From Tchaikovsky's Guide to the Practical Study of Harmony [33], motifs or sequences of notes capturing an idea are important for sound composition. Consistency within three or more distinct notes, repeated throughout the piece are positively rewarded.

### 3.3. Software Design
### 3.3a Deep Learning

The featured data vector in this neural network is referred to as the 'Note State Matrix' shown in Figure 1. This represents the 'play' and 'articulate' state of each note over the range of Midi values and for each time step over a specified period of time (i.e. 8 measures at 16 time steps per measure). The model takes as input a single batch of these feature data vectors, a single 4D tensor referred to as the 'Note_State_Batch'. The original raw musical data in the form of .MIDI files are first preprocessed to generate each Note_State_Batch using the Python-Midi package extracted from [42]. In this work, I used this package only to import MIDI file segments as Note_State_Batches, as well as to create MIDI files from the Generated Samples. It may be of interest in further work to enhance the processed feature data vectors to include other musical features such as volume.

$$\text{Note State Matrix} = \begin{bmatrix} [p,a]_N^{(1)} & \cdots & [p,a]_N^{(T)} \\ & \vdots & \\ [p,a]_1^{(1)} & \cdots & [p,a]_1^{(T)} \end{bmatrix} \text{ a batch of which constitute a 'Note\_State\_Batch.'}$$

Fig. 1: Note State Matrix is the processed, feature vector for the neural network. N is the # Midi note states extracted from the songs, T is the # time steps in the batch, 'p' indicates the binary value of the note being played, and 'a' indicates the binary value of the corresponding articulation.

The overall structure of the code was broken into two main tasks: training/validating the model numerically and then using the trained model to generate new .MIDI files for qualitative evaluation. Both functions use the same Model Graph in different contexts: The training task, shown at a high level in Figure 2, iteratively inputs a Note_State_Batch into the model, runs the model through all of the corresponding time steps and notes present in the batch, and then outputs a tensor of corresponding 'Logits', or inverse sigmoid probability that a given note at a given time step is played/articulated. The log likelihood of the input data is interpreted as the ability of the model to take as input a vector of notes at a given time step and to predict the set of notes at the subsequent time step. The Loss Function, pseudo-code of which is shown in Figure 3, calculates cross-entropy between the generated Logits and the Note_State_Batch (after lining up the Logits to the Note_State_Batch elements corresponding to one time step in the future).

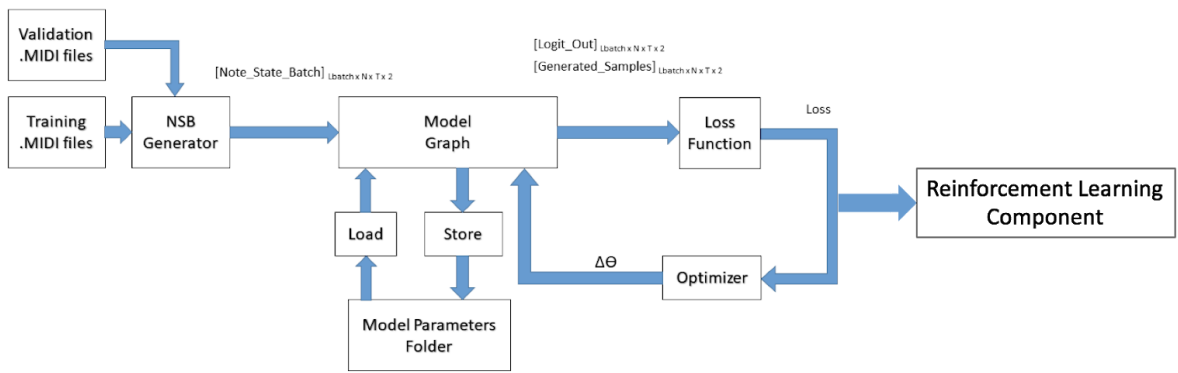

Fig. 2: High-level view of the Training Graph

**Arguments:**

- [Logits] $L_{batch} \times N \times T \times 2$ (inverse sigmoid of Probability that play/articulate = 1)
- Labels(t-1) = [Note_State_Batch] $L_{batch} \times N \times T \times 2$

$$\text{cross\_entropy} = \text{sigmoid\_cross\_entropy\_with\_logits}(\text{logits}=\text{logits}, \text{labels}=\text{Labels})$$
$$= -\text{note\_state} \ln(\sigma(\text{logits})) + -(1 - \text{note\_state}) \ln(1 - \sigma(\text{logits}))$$
$$= -\text{note\_state} \ln(\text{Probability}=1) + -(1 - \text{note\_state}) \ln(\text{Probability}=0)$$

$$\text{Loss} = \frac{1}{TNL_{batch}} \sum_{b=1}^{L_{batch}} \sum_{t=1}^{T} \sum_{n=1}^{N} \text{cross\_entropy}$$

$$\text{Log-likelihood at 1 time step} = -\frac{1}{TL_{batch}} \sum_{b=1}^{L_{batch}} \sum_{t=1}^{T} \sum_{n=1}^{N} \text{cross\_entropy}$$

**Returns:**

- Loss (scalar)

Fig. 3: Pseudo-code for Loss definition and calculation

During the music generation task, presented in Fig. 4, the model is iteratively run through one time step, every time feeding back the Generated Samples as the Note_State_Batch input for the subsequent time step. This samples are accumulate, and this produces a tensor of Generated Samples in the form of the Note_State_Batchof arbitrary time length. The Generated Samples are then converted to .MIDI files using post-processing functions from [42] for qualitative evaluation.

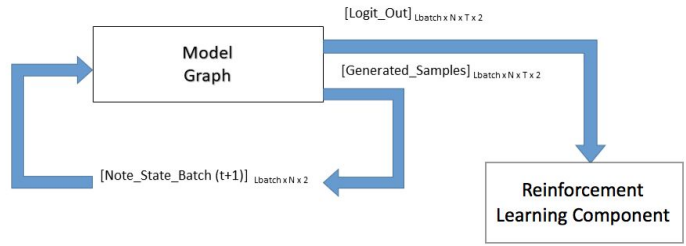

Fig. 4: High-level view of music generation graph.

A functional breakdown of the Model Graph, itself, is shown in Figure 5. Pseudo code of the first stage of the model, referred to as the 'Input Kernel'', is shown in Figure 6. The Input Kernel takes a Note_State_Batch as its input and for each note/articulation pair, generates an expanded vector that consists of: 1) the Midi note number, 2) a one hot vector of the note's pitch class, 3) window of the play/articulation values relative to the 'n'th note (the effective convolutional kernel aspect of the model), 4) a vector of the sum of all played notes in each pitch class, and 5) a binary-valued vector representing the 16 value position of the note within a measure.

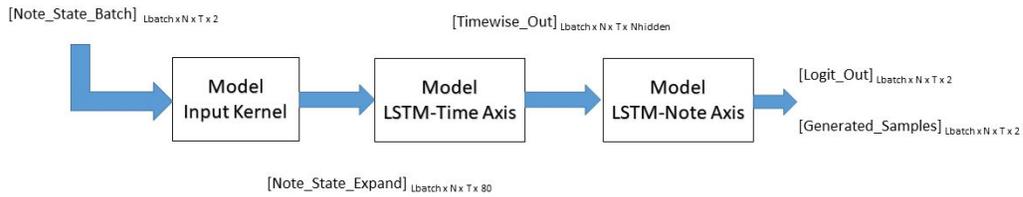

Fig. 5: Breakdown of Model Graph

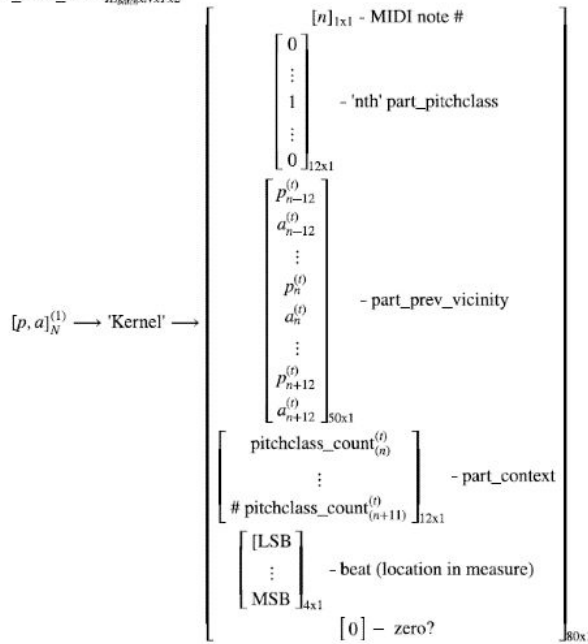

Fig. 6: Pseudo code for Input Kernel. This code is performed in parallel note-wise, time-wise, and sample-wise.

The second stage is referred to as the Timewise LSTM stage, pseudo-code for which is shown in Figure 7. In this block, an LSTM cell is run along the time axis for the length of the batch time dimension. This operation is performed on the Note_State_Expand vector for every note in parallel with tied weights. This part of the graph captures the sequential patterns of the music and, in combination with the Input Kernel, preserves translation invariance due to the input window of relative notes and the tied LSTM weights across all notes. Due to these tied weights, the computations can be run in parallel across notes and across Note State Matrix samples as separate effective batches. The only required sequential aspect is along the time axis. An arbitrary number of cascaded LSTM cells can be run, and a dropout mask is applied after each cell.

Arguments:
- [Note_State_Expand] $L_{batch} x N x T x L_{filt}$ where $L_{filt}$ = 80

$$t = 1 : T$$
$$h_{(1)n}^{(t)} = \text{LSTM}(h_{(1)n}^{(t-1)}, \text{Note\_State\_Expand}_n^{(t)})$$
$$h_{(2)n}^{(t)} = \text{LSTM}(h_{(2)n}^{(t-1)}, h_{(1)n}^{(t)})$$
$$\vdots$$
$$\text{timewise\_out} = h_{(num-t-layer)n}^{(t)} = \text{LSTM}(h_{(num-t-layer)n}^{(t-1)}, h_{(num-t-layer-1)n}^{(t)})$$

Returns:
- [timewise_out] $L_{batch} x N x T x N_{h_{num-t-layer}}$

Fig. 7: Pseudo code for Timewise LSTM stage. This code is performed in parallel note-wise and sample-wise.

The final stage in the Model Graph as described in the block diagram is the Notewise LSTM stage, pseudo-code for which is shown in Figure 8. This is a potentially one or multi-layered LSTM stage like the Timewise LSTM, also with dropout after each layer. However, instead of running sequentially along the time axis, this stage runs sequentially along the note axis. Furthermore, this section includes the 'local' feedback of generated samples into its input. After each 'note step', the LSTM cell produces a pair of logits representing the inverse sigmoid of the probability of generating a play/articulation for that note. Next, a play and articulation sample are drawn from this Bernoulli distribution. If the play sample is a '0' for 'not played', the articulation sample is forced to '0', as well, to avoid the generation of any values not present in the input data. The generated sampled pair at note (n-1), concatenated with the input of the timewise LSTM stage at note (n), is fed back into the input of the notewise LSTM for step (n). This feedback creates a conditional probability for each note based on the actual values generated for lower notes. This helps prevent dissonant simultaneous notes from being played. The final output tensors of the Model Graph are the batch of Logits and corresponding Generated Samples that are used for training and music generation, respectively.

Arguments
- [timewise_out] $L_{batch} \times N \times T \times N_{h_{num-t-layer}}$

for n = 1 to N:

$$\text{cell\_input}_n = \begin{bmatrix} \text{timewise\_out}_n \\ \text{note\_gen}_{n-1} \end{bmatrix}$$

$$h_{(1)_n} = \text{LSTM}(h_{(1)_{(n-1)}}, \text{cell\_input}_n)$$

$$h_{(2)_n} = \text{LSTM}(h_{(2)_{(n-1)}}, h_{(1)_n})$$

$$\vdots$$

$$h_{(\text{num-n-layer})_n} = \text{LSTM}(h_{(\text{num-n-layer})_{(n-1)}}, h_{(\text{num-n-layer-1})_n})$$

$$\text{Logits}_n = W h_{(\text{num-n-layer})_n} + b = \begin{bmatrix} \text{Logits}_n - play \\ \text{Logits}_n - artic \end{bmatrix}$$

$$\text{note\_gen}_n = \text{Sample}[\sigma(\text{Logits}_n) = \text{Prob}(\text{note\_n} = 1)] = \begin{bmatrix} \text{play\_gen}_n \\ \text{artic\_gen}_n \end{bmatrix}$$

if (play_gen$_n$ = 0) then artic_gen$_n$ = 0

Return:
- [Logits]$_{L_{batch} \times N \times T \times 2}$
- [note_gen]$_{L_{batch} \times N \times T \times 2}$

Fig. 8: Pseudo Code for Notewise LSTM Stage. This code is performed in parallel time-wise and sample-wise.

### 3.3b Reinforcement Learning

To capture the musical rules described in the last paragraph of section 3.2, the musical characteristics present in the MIDI file were used. Taking a cue from Natasha Jacques' characterization, three octaves of pitches starting from MIDI pitch forty eight are encoded as: two = C2, three =C#3, four = D3, etc, thirty-seven = B5. For instance, the sequence {3, 1, 0,1} encodes an eighth note with pitch C sharp, followed by an eighth note rest. The sequence {2, 4, 6, 7} encodes a melody of four sixteenth notes: C3, D3, E3, F3. A length 38 one-hot encoding of these values is used for both network input and network output.

The learned weights of the Biaxial Model were used to prime the three sub-networks in the RL model. The overview of the RL model is shown in Figure 9.

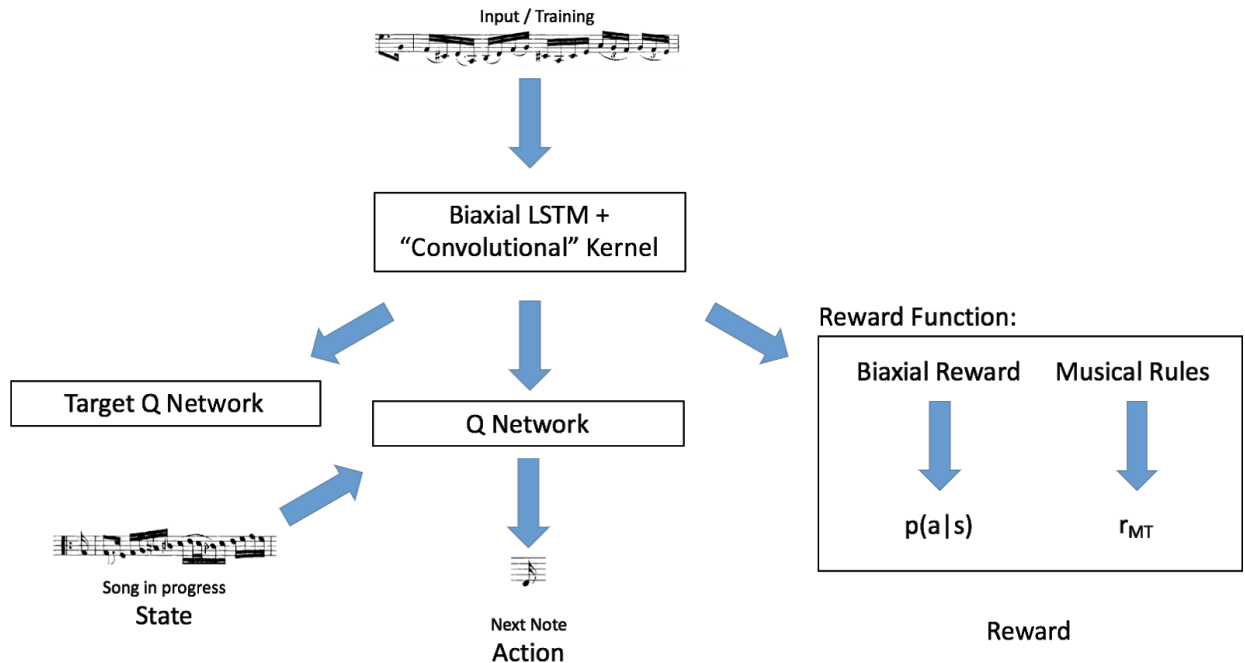

Fig. 9: The Biaxial model is trained, and initializes the target Q network, Q network, and the Biaxial Reward.

This paper's RL model was trained for 50,000 iterations. For consistency with the Biaxial model, the optimizer used was ADADELTA. The learning rate selected was 1.0. Batch size of 32 was used. In Jacques' model, the optimizer used was Adam [43]. Gradients were clipped and the the learning rate was set to .5, with a momentum of .85 to decay the learning rate every 1000 steps. Jacques' model was trained for 1,000,000 iterations. Due to time and computational resource constraints, I elected not to use Jacques' model specifications for the number of iterations. The RL model used in this paper did take a cue from Jacques' for other aspects of the RL model: the reward discount factor of $\gamma=.5$ was used. The Target Q-network's weights $\theta$ were updated incrementally to be effectively similar to the Q-network's $\theta$ according to the formula $(1 - \eta)\theta^{-} + \eta\theta$, where $\eta = .01$ is the Target-Q-network update rate. Once again, due to limited time and computational resources a minimum number of experiments were run. The modifying constant of the musical theory reward $r_{MT}$, c, was set to equal .5. Boltzmann exploration and epsilon-greedy were coded, but only epsilon-greedy was run. Additionally, a Q-learning method with eligibility traces was coded but due to limited time and computational resources was not run. The RL model used Q-learning with the policy from the Biaxial model used as the cross entropy reward. Next section will contain results.

## 4. Results
### 4.1. Quantitative Analysis

Table 1 shows the log-likelihood performance of Daniel Johnson's models, as well as that of the pure Biaxial LSTM model implemented in this work. The results obtained in this work were, in general, on par with the survey of models reported by the Daniel Johnson but somewhat inferior to the corresponding model. However, limitation of training time (due to time and resource constraints) most likely plays the largest role in this discrepancy. In his blog, Daniel Johnson estimated roughly 24-48 hours of training to capture the quality of his music samples, whereas the training for this work consisted of about 13 hours on a limited set of 1-2 dozen of Bach's fugues from [36]. In addition, the author's paper reported optimization using RMSprop whereas his blog, which seemed to represent the latest of his progress, reported Adadelta. This work started with the latter, but more experimentation needs to be done to fine tune such hyper parameters.

| Model | Log-Likelihood | Hours Trained |
|---|---|---|
| Random | -61 | -- |
| TP-LSTM-NADE | -5.44, -5.49 | 24 - 48 |
| BALSTM | -4.90, -5.00 | 24 - 48 |
| BALSTM (this work) | -6.27, -7.93 (test) -5.16, -6.59 (train) | 16 |

Table 1: The top 3 rows represent the Log-likelihood performance reported by the original author for random weighted, LSTM-NADE, and Bi-axial LSTM networks tested on the Piano-Midi.de data set. The two values represent the best and median performance across 5 trials. For the bottom row, the two values represent best and median across 100 trials for the BALSTM in this work, scaled by 88/78 to normalize it to the number of notes used by Daniel Johnson.

Table 2 shows the performance of the models on the hard-coded music theory rules. Results of this paper's Biaxial model are in column 2 while Natasha Jacques' priming model called Note RNN are in column 3. In column 4, is this paper's RL component primed with the Biaxial model while column 5 is Natasha Jaques' complete RL model. The statistics were computed by randomly generating 1,000 songs from each model.

The results show that RL helps improve almost in every column in the metrics column. Of note is the Biaxial model outperforms the Note RNN model in almost every metric that requires context: repeated notes and nearly all the auto-correlation measures. This follows from the addition of the convolutional-esque kernel which slides over notes providing greater links between notes. This follows from the fact that convolution and autocorrelation are operations with minor differences. In terms of the addition of the RL component,

This paper's RL model and Jacques' RL model perform in a pattern similar to the pure deep learning model. This paper's RL model outperforms in contextual operations, like autocorrelation and is outperformed in every other musical theory category. This follows again from the incorporation of a convolutional operation. The outperformance follows from the significant gap in training iterations: 1,000,000 iterations for Natasha Jacques' model and 50,000 iterations for this paper's model. Given the discrepancy in training time, the incorporation of the convolutional operation indicates a significant increase in performance and a fruitful addition to the priming model. The addition of RL clearly indicates an improvement: learning play notes in key, autocorrelation decreases, more unique high and low notes, fewer leaps, and more motifs.
In terms of the numbers themselves, the increase in performance on the metrics is tied to the strength of the reward signal for the specific behavior. For instance, playing a note frequently received a serious negative reward. Rewards can be adjusted to alter performance on desired metrics.

| Metric | Biaxial Model (this paper) | Note RNN (benchmark) | Q (this paper) | Q (benchmark) |
|---|---|---|---|---|
| Notes repeated | **47.7%** | 63.3% | 17.0% | **0.0%** |
| Mean Autocorrelation – lag 1 | **0.13** | -.16 | **-0.05** | -.11 |
| Mean Autocorrelation – lag 2 | -0.23 | **.14** | **0.09** | .03 |
| Mean Autocorrelation – lag 3 | **-0.12** | -.13 | **0.04** | .03 |
| Notes not in key | 0.4% | **.1%** | 2.7% | **1.00%** |
| Melody starting with tonic | 0.1% | **.9%** | 19.2% | **28.8%** |
| Leaps resolved | 69.1% | **77.2%** | 79.3% | **91.1%** |
| Melodies with unique highest note | 42.9% | **64.7%** | 34.7% | **56.4%** |

| Melodies unique lowest note | **55.7%** | 49.4% | 48.9% | **51.9%** |
| --- | --- | --- | --- | --- |
| Notes in motif | 3.2% | **5.9%** | 49.3% | **75.7%** |
| Notes in repeated motif | **.04%** | 0.007% | .07% | **.11%** |

Table 2- Comparison of model performance on music theory rules. Rows starting with "Notes not in Key" and above (inclusive), lower percentages are better. Rows starting with "Melody starting with tonic and below (inclusive), higher percentages are better.

### 4.2. Qualitative Analysis

To go beyond my untrained ear, I asked Professor Jospeh Dubiel of Columbia University, former Chair of the Music department and the Chair of the Music Theory Area Committee, for his thoughts on the piece. Professor Dubiel generously answered in detail. We communicated over email, so I will quote him fully.

> "Generated" can mean a lot of things, especially in computer music. Sometimes it refers to the synthesis of the sounds themselves, in contrast to what I assume in happening in your piece, sampling from a bank of prerecorded "piano" sounds. What really put me in a questioning mood was the sense of the sample you sent as losing its way every few beats: cohering for a very short time--from a few beats to a few seconds--then making an apparently pointless change, to something that might have a local connection, but that is substantially different in direction (when there's direction at all), pacing, and sometimes even style. This unevenness of continuity made me wonder whether the unit of selection was a single note, or something longer, perhaps a short figure drawn from a preexistent repertoire of such figures, or modeled on figures in such a repertoire.

> Your sample doesn't quite sound like an intentional jump-cut piece, but might come off a little more like human beginner's effort to make one of those than like a continuous composition in a traditional sense. Listening to it one more time, I realize that I may have exaggerated the sense of cutting in order to try to bring the sample close to *some* kind of actual music. If I do my best to listen to it as a single succession, it's the *rhythmic* discontinuity, the constant stopping and starting, that makes this unbelievable. The pauses just come whenever, not when any motivic, harmonic, or phraseological action seems to have been completed (or at moments that would make sense as dramatic interruptions). The final stop is definitely one of these: I come to the end of the sample with absolutely no sense that it was a composition, as opposed to an arbitrary selection from a stream of indefinite length. I can imagine how this might happen if the models used to produce it are as narrowly focused on pitch as these may have been."

Qualitatively, the samples produced by this model when listened to by a professional clearly suggest limitations. The music breaks down in its ability to create clear transitions between larger ideas in the piece as a whole. There is no deeper structure. The sample also makes poor use of negative space, few pauses are present in the work. Due to the lack of global structure, the music has a mechanical feel. An important note is the length of training time. When the model is trained for 30 minutes, the music generated is sparse and significantly less consistent and coherent. When trained for 2 hours, the difference is dramatic.  Clear relationships between generated music and the corresponding training files developed.

**4.3. Discussion of Insights Gained**

It became clear how the variability and complexity of music on which the model was trained affected the outcome.  Training a newly initialized model on a large data set consisting of significant variability in music segments (i.e. fast monodic and slow polyphonic) tended to create a model that seemed to be confused at first.  Trying to learn such a range of features requires a complex function needing long training times.  Training on a set consisting of 22 of Bach's fugues from [36] obtained better results more quickly than training on the 120 Piano-de-Midi for modest training times < 2 hours. However, it became evident that very long training times were required, in general to produce decent music. It was clear the music was gradually learning rhythm and chord structures, however it sounded as if a human were learning to play piano by trying to play songs that were too difficult. One possible training strategy may be to train on a succession of increasingly difficult songs, graduating the model manually, or perhaps in an automated fashion once a certain ability or log likelihood was achieved. In addition to songs of different 'level-of-difficulty', training could begin on very short time segments and increase to very long segments to allow the model to learn basic structure in addition to longer musical form. This approach to learning is in the vein of curriculum learning [44].

In terms of future work, it would be fruitful to add to the bi-axial LSTM a component that focused on structure alone. There has been good work showing the merits of using Restricted Boltzmann Machines to model chord progressions and other forms of harmonic and melodic structure. Additionally, an effective model could incorporate genetic algorithms. The line of thinking would be to train the model on some simple music and set the fitness score as a proxy for novelty, and allow the algorithm to generate mutations to add complexity to the piece over time. Other dynamic optimization techniques like ant colony optimization could also prove to be effective. Another model design that would be effective would be Generative Adversarial Networks (GANs) [45] which have achieved remarkable progress in generating photo-realistic images and as such should provide effective musical generation. Moving beyond the deep learning priming model, a more innovative approach is to rely on reinforcement learning and incorporate a more refined sense of exploration in the music generation. A potential refinement can enter in the sampling the action space: paths or musical measures explored may have different rewards associated with the distribution. In a similar vein, I suggest exploration can enter by building into the model a sense of choosing the action that maximizes the expected reward with respect to a randomly drawn belief. This method is called Thompson Sampling [46] and draws from the Multi-armed bandit literature. Another idea is to move beyond the one-hot encoding of notes and look at a vectorized representation musical chord embeddings, akin to word2vec for music. There has been work that creates a chord2vec tool [47], which can improve performance. Further, incorporating a hierarchical training strategy could prove to be effective: taking an idea from the image segmentation literature, and separate out different instruments and have different models focus on representing with fidelity the sound of say the cello. This could improve performance and sound quality in the produced samples. Additionally, an issue with creating global coherence in the music is having the algorithm learn when to shift between ideas in the composition. Having clear labels for transitions between motifs via a spatio-temporal (note axis and time axis) labeling mechanism could improve global coherence. Additionally, currently the music is formulated as a discrete time problem: the song is sliced into discrete chunks and solved recursively via Q learning through Bellman's Equation. There has been work to create continuous deep Q learning with model based acceleration, which would limit loss of information in connecting the local subproblems, or linking the discrete notes [48]. The Hamilton Jacobi Bellman equation could prove to be useful. Computational time and resources scale at an unpalatable rate when using more and more complex models, called the "Curse of Dimensionality". Working in the spectral domain, affords potentially significant increases in speed. This seems to be a natural extension given the raw waveform of the file [49]. There is much theoretical work that can be done to link the process of exploring the action space to stochastic optimal control [50]. This can be extended with the Fenyman-Kac formula [49] which offers a link between partial differential equations and stochastic processes. The gist of the idea is to consider the process of music selection as an interacting particle system. This affords a nice formulation in terms of game theory: mean field games [50] and randomized equilibria [51] map on nicely to selecting different sub-equilibria of high reward actions. Convergence and stability can be shown in discrete time by solving the Hamiltonian, making a fixed point argument and finding an equilibrium concept. Beyond the reinforcement learning, further work can be done on the musical side by incorporating expert rules for musical creation.

## 5. Conclusion

In this paper, a two layer LSTM model with a convolutional-esque kernel capable of learning harmonic and melodic rhythmic probabilities from polyphonic MIDI files of Bach was created. The model design was explained, with an eye to key functional principles of flexibility and generalizability. The model was extended with a reinforcement learning (DQN) approach. The underlying logic and method of training and generation of algorithmic music were presented. Further, the outputs of the model were analyzed in a quantitative and qualitative fashion. Some suggestions were then put forward for future work.

## 6. Acknowledgement